\documentclass[12pt]{article}
\usepackage{amssymb,amscd,array}
\catcode `\@=11
\@addtoreset{equation}{section}

\def\harr#1#2{\smash{\mathop{\hbox to .5in{\rightarrowfill}}
\limits^{\scriptstyle#1}_{\scriptstyle#2}}}

\def\harrl#1#2{\smash{\mathop{\hbox to .5in{\leftarrowfill}}
\limits^{\scriptstyle#1}_{\scriptstyle#2}}}

\newcommand{\be}{\begin{equation}}
\newcommand{\ee}{\end{equation}}
\newcommand{\bea}{\begin{eqnarray}}
\newcommand{\eea}{\end{eqnarray}}

\newtheorem{thm}{Theorem}[section]
\newtheorem{rem}[thm]{Remark}

\newtheorem{prop}[thm]{Proposition}
\begin{document}
\begin{titlepage}
\begin{center}
{\bf \Large{Massive Supersymmetric Quantum Gauge  Theory \\}}
\end{center}
\vskip 1.0truecm
\centerline{Dan Radu Grigore\footnote{e-mail: grigore@theory.nipne.ro}}
\centerline{Dept. Theor. Phys., Inst. Atomic Phys.,
Bucharest-M\u agurele, MG 6, Rom\^ania}
\vskip5mm
\centerline{Markus Gut and G\"unter Scharf
\footnote{e-mail: scharf@physik.unizh.ch}}
\vskip5mm
\centerline{Institute of Theor. Phys., University of Z\"urich}
\centerline{Winterthurerstr. 190, Z\"urich CH-8057, Switzerland}
\vskip 2cm
\bigskip \nopagebreak
\begin{abstract}
\noindent
We continue the study of the supersymmetric vector multiplet in a purely 
quantum framework. We obtain some new results which make the connection
with the standard literature. First we construct the one-dimensional
physical Hilbert space taking into account the (quantum) gauge structure 
of the model. Then we impose the condition of positivity for the scalar
product only on the physical Hilbert space. Finally we obtain a full
supersymmetric coupling which is gauge invariant in the supersymmetric
sense in the first order of perturbation theory. By integrating out the
Grassmann variables we get an interacting Lagrangian for a massive 
Yang-Mills theory related to ordinary gauge theory; however the
number of ghost fields is doubled so we do not obtain the same
ghost couplings as in the standard model Lagrangian.
\end{abstract}

\newpage\setcounter{page}1
\end{titlepage}
\section{Introduction}

The supersymmetric gauge theories are constructed using the
vector supersymmetric multiplet \cite{FP} (see also \cite{WB}, \cite{Wes} 
\cite{Wei}, \cite{GGRS}, \cite{Bi}, \cite{Fi}, \cite{Li}, \cite{So}, 
\cite{Pi1}, \cite{Pi2}, \cite{Sa},  etc.) This is a superfield for a
$N = 1$
supersymmetric theory which describes only particles of spin
$s \leq 1$
and it contains in particular exactly one particle of spin $1$ in the
one-particle Hilbert space. In the standard literature quoted above this
model is considered mainly as a classical field theory and the quantization
is performed by path integral method. We have considered in detail the vector
multiplet in a previous paper \cite{GS2} in a pure quantum framework. 
We have analyzed there the question of positivity for the whole multiplet
and have obtained some restrictions on the free parameters of the model
(the mass and some free parameters appearing in the causal commutator).
Then we have analyzed the gauge structure associated to the vector
multiplet (constructing some natural ghost fields which are chiral fields)
and tried to construct an interaction Lagrangian built only from superfields 
and gauge invariant in a supersymmetric sense. One can obtain non-trivial
solutions for this problem but it is not so easy to find a solution
which gives, after integrating the Grassmann variables, the usual interaction
for a system of massive Yang-Mills fields. In fact, our ansatz for the
coupling in \cite{GS2} was not general enough for this goal; this shortcoming 
will be removed here.

In this paper we consider only vector multiplets of non-null mass and we
use the gauge structure already determined in \cite{GS2}. Using the
explicit expression of the gauge charge $Q$ (we remind that $Q$ is 
an operator acting in a Hilbert space
${\cal H}$
with a non-degenerate sesqui-linear form
$<\cdot, \cdot>$
verifying
$Q^{2} = 0$
such that the physical Hilbert space is
${\cal H}_{phys} \equiv Ker(Q)/Im(Q)$)
we are able to determine the structure of the one-dimensional physical 
Hilbert space
$Ker(Q)/Im(Q) \cap {\cal H}^{(1)}$;
then we impose the condition of positivity of the scalar product only on this
sub-space. We obtain weaker conditions on the parameters of the model
than in our previous work \cite{GS2}. This is done in Section \ref{sg}.
Then we make a more general ansatz for the supersymmetric interaction 
Lagrangian and we succeed in finding a solution 
having all the required properties. We require that the corresponding 
interaction Lagrangian 
$t(x)$
should be renormalizabe (so it must have canonical dimension 
$\leq 4$)
and it should follow by integrating out the Grassmann variables 
\be 
t(x) \equiv \int d\theta^{2} d\bar{\theta}^{2} T(x,\theta,\bar{\theta});
\label{tT}
\ee
moreover we require that
$T(x,\theta,\bar{\theta})$
is gauge invariant in the supersymmetric sense:
\bea
[Q, T(x,\theta,\bar{\theta}) ] = 
{\cal D}^{a} T_{a}(x,\theta,\bar{\theta}) +
\bar{\cal D}_{\bar{a}} \bar{T}^{\bar{a}}(x,\theta,\bar{\theta})
\label{gauge-susy}
\eea
where
\bea
{\cal D}_{a} \equiv {\partial \over \partial \theta^{a}} 
- i \sigma^{\mu}_{a\bar{b}} \bar{\theta}^{\bar{b}} \partial_{\mu}
\qquad
\bar{\cal D}_{\bar{a}} \equiv 
- {\partial \over \partial \bar{\theta}^{\bar{a}}}
+ i \sigma^{\mu}_{b\bar{a}} \theta^{b} \partial_{\mu};
\label{calD}
\eea
we call {\it SUSY-divergences} the expressions appearing in the right hand 
side of formula (\ref{gauge-susy}). {\it A priori} one should include
in the right hand side of (\ref{gauge-susy}) space-time total divergence
$\partial_{\mu}T^{\mu}$
also. But if we use the identity
\be
\{ {\cal D}_{a}, \bar{\cal D}_{\bar{b}} \} = 
- 2 i \sigma^{\mu}_{a\bar{b}}~\partial_{\mu}
\ee
we can eliminate the space-time divergence by redefining the expressions
$T_{a}$
and
$\bar{T}^{\bar{a}}$.

We get a possible interaction for a system of massive Yang-Mills fields
verifying (\ref{tT}) and (\ref{gauge-susy}). Because the super-ghost
fields are chiral we obtain a model with twice as many ghost fields as
the usual standard model. As a consequence the resulting coupling is
different from that in the standard model. Whether this has implications
for phenomenology remains to be investigated.

\section{The Gauge Structure of the Vector Multiplet \label{sg}}

We use the notations and definitions from \cite{GS2}. The {\it vector
multiplet} is by definition given by the expression
\bea
V(x,\theta,\bar{\theta}) = C(x) + \theta\chi(x) + \bar{\theta} \bar{\chi}(x) 
+ \theta^{2}~\phi(x) + \bar{\theta}^{2}~ \phi^{\dagger}(x)
\nonumber \\
+ (\theta \sigma^{\mu} \bar{\theta})~v_{\mu}(x)
+ \theta^{2}~\bar{\theta}\bar{\lambda}(x) 
+ \bar{\theta}^{2}~\theta\lambda(x) 
+ \theta^{2} \bar{\theta}^{2}~ d(x)
\label{V}
\eea
with the reality condition:
\be
V^{\dagger} = V;
\ee
here
$C$ and $d$ are real scalar fields,
$\phi$ is a complex scalar field,
$v_{\mu}$
is a hermitian vector field,
$\chi_{a}$
and
$\lambda_{a}$
are Dirac spinor fields and
$\theta_{a}$
are Grassmann parameters. The major difference with respect to \cite{GS2}
is that we do not suppose that the Hilbert space in which these fields
live has a positive defined scalar product; we suppose that we have only 
a indefinite sesqui-linear form
$<\cdot, \cdot>$.
The construction of the fields is done in the same way as in the usual case
by applying these fields on the vacuum state
$\Omega$,
because Borchers algebra can be constructed regardless of the 
positive-defined character of the sesqui-linear form. We also 
suppose that all these fields are of mass $m$:
\be
(\partial^{2} + m^{2}) V = 0
\label{kg}
\ee
and we take $m$ to be strictly positive.

We remind from \cite{GS2} the decomposition of $V$ which will be very useful
in the following: 
\begin{prop}
The vector superfield $V$ can be written as follows
\be
V = \sum_{j=0}^{2} P_{j} V
\ee
where the expressions
$P_{j}, \quad j = 0,1,2$
are given by
\be
P_{0} \equiv - {1\over 8m^{2}} {\cal D}^{a} \bar{\cal D}^{2} {\cal D}_{a}, 
\quad
P_{1} \equiv {1\over 16m^{2}} {\cal D}^{2} \bar{\cal D}^{2}, \quad
P_{2} \equiv {1\over 16m^{2}} \bar{\cal D}^{2} {\cal D}^{2}.
\label{projectors}
\ee

The expressions
$P_{j}, \quad j = 0,1,2$
are projectors on the mass shell i.e. they verify
\bea
P_{j} P_{k} = 0, \quad \forall j \not= k,
\nonumber \\
P_{j}^{2}~V = P_{j}~V, \quad \forall j.
\eea

The components
$V_{j} \equiv P_{j}~V, \quad j =1,2$
of $V$ verify
\be
{\cal D}_{a}~V_{1} = 0, \quad \bar{\cal D}_{a}~V_{2} = 0.
\label{chiral-dec}
\ee
\label{decomp-v}
\end{prop}

The three components are called {\it chiral, anti-chiral} and 
{\it transversal} components of $V$, respectively. The appellative
transversal is justified by the explicit formula:
\bea
V_{0} = - {2\over m^{2}}~D^{\prime}
\nonumber \\
D^{\prime} = d^{\prime} - 
{i\over 2}~\theta \sigma^{\mu} \partial_{\mu}\bar{\lambda}^{\prime}
+ {i\over 2}~\partial_{\mu}\lambda^{\prime} \sigma^{\mu} \bar{\theta}
- {1\over 2}~(\theta \sigma^{\mu} \bar{\theta})~( m^{2}~g_{\mu\rho}
+ \partial_{\mu}\partial_{\rho} )v^{\rho}
\nonumber \\
- {m^{2} \over 4}~\left(\theta^{2}~\bar{\theta}\bar{\lambda}^{\prime}
+ \bar{\theta}^{2}~\theta\lambda^{\prime} \right)
- {m^{2} \over 4}~\theta^{2} \bar{\theta}^{2}~d'
\label{d-prime}
\eea
where
\bea
\lambda^{\prime}_{a} \equiv \lambda_{a} + {i\over 2} \sigma^{\mu}_{a\bar{b}} 
\partial_{\mu}\bar{\chi}^{\bar{b}}
\nonumber \\
d^{\prime} \equiv d - {m^{2} \over 4} C;
\eea
in particular the superfield
$V_{0}$
contains only one Majorana spinor
$\lambda^{\prime}$,
one scalar field
$d^{\prime}$
and a real vector field
\be
v^{\prime}_{\mu} \equiv 
\left(g_{\mu\rho} + {1\over m^{2}} \partial_{\mu}\partial_{\rho}\right)v^{\rho}
\ee
verifying the transversality condition
\be
\partial^{\mu}v^{\prime}_{\mu} = 0.
\ee

Next we remind the generic form of the causal commutator of the
vector superfield. Let us consider the causal commutator
\be
~[ V(X_{1}), V(X_{2}) ] = - i D(X_{1};X_{2})~ {\bf 1};
\ee
then we have according to theorem 3.7 of \cite{GS2}
\be
D(X_{1};X_{2}) = \sum_{j=1}^{4} c_{j}~D_{j}(X_{1};X_{2})
\label{ddd}
\ee
where
\bea
D_{1}(X_{1};X_{2}) = \exp [i\left( \theta_{1} \sigma^{\mu} \bar{\theta}_{2}
- \theta_{2} \sigma^{\mu} \bar{\theta}_{1} \right) \partial_{\mu} ]~
D_{m}(x_{1}-x_{2})
\nonumber \\
D_{2}(X_{1};X_{2}) = 
(\theta_{1}-\theta_{2})^{2} (\bar{\theta}_{1}-\bar{\theta}_{2})^{2}~
D_{m}(x_{1}-x_{2})
\nonumber \\
D_{3}(X_{1};X_{2}) = \exp [i\left( \theta_{1} \sigma^{\mu} \bar{\theta}_{2}
- \theta_{2} \sigma^{\mu} \bar{\theta}_{1} \right) \partial_{\mu} ]
[ (\theta_{1}-\theta_{2})^{2}  + (\bar{\theta}_{1}-\bar{\theta}_{2})^{2} ]~
D_{m}(x_{1}-x_{2})
\nonumber \\
D_{4}(X_{1};X_{2}) = i~\exp [i\left( \theta_{1} \sigma^{\mu} \bar{\theta}_{2}
- \theta_{2} \sigma^{\mu} \bar{\theta}_{1} \right) \partial_{\mu} ]
[ (\theta_{1}-\theta_{2})^{2} - (\bar{\theta}_{1}-\bar{\theta}_{2})^{2} ]~
D_{m}(x_{1}-x_{2}).
\label{basis}
\eea

It is useful to write down the corresponding causal (anti)commutation relations
of the component fields. A straightforward computation leads to:
\bea
~[ C(x), C(y) ] = - i~c_{1}~D_{m}(x-y)
\nonumber \\
~[ C(x), d(y) ] = - i~c_{2}~D_{m}(x-y)
\nonumber \\
~[ C(x), \phi(y) ] = - i~(c_{4} - i c_{3})~D_{m}(x-y)
\nonumber \\
~[ \phi(x), \phi^{\dagger}(y) ] = 
- i~\left( {m^{2}\over 4}~c_{1} + c_{2}\right)~D_{m}(x-y)
\nonumber \\
~[ \phi(x), d(y) ] = {m^{2}\over 4}~(c_{4} - i c_{3})~D_{m}(x-y)
\nonumber \\
~[ \phi(x), v_{\mu}(y) ] = (c_{3} + i c_{4})~\partial_{\mu}D_{m}(x-y)
\nonumber \\
~[ d(x), d(y) ] = - {im^{4}\over 16}~c_{1}~D_{m}(x-y)
\nonumber \\
~[ v_{\mu}(x), v_{\rho}(y) ] = 
i~c_{1}~\partial_{\mu}\partial_{\rho}~D_{m}(x-y)
+ i~\left({m^{2}\over 2}~c_{1} - 2 c_{2}\right)~g_{\mu\rho}~D_{m}(x-y)
\nonumber \\
\left\{ \chi_{a}(x), \chi_{b}(y) \right\} 
= 2 (c_{4} - i c_{3})~~\epsilon_{ab}~D_{m}(x-y),
\nonumber \\
\left\{ \chi_{a}(x), \bar{\chi}_{\bar{b}}(y) \right\} =
c_{1}~\sigma^{\mu}_{a\bar{b}}~\partial_{\mu}D_{m}(x-y)
\nonumber \\
\left\{ \lambda_{a}(x), \lambda_{b}(y) \right\} 
= - {m^{2}\over 2}(c_{4} - i c_{3})~\epsilon_{ab}~D_{m}(x-y),
\nonumber \\
\left\{ \lambda_{a}(x), \bar{\lambda}_{\bar{b}}(y) \right\} =
{m^{2}\over 4}~c_{1}~\sigma^{\mu}_{a\bar{b}}~\partial_{\mu}D_{m}(x-y)
\nonumber \\
\left\{ \chi_{a}(x), \lambda_{b}(y) \right\} 
= - 2i~c_{2}~\epsilon_{ab}~D_{m}(x-y),
\nonumber \\
\left\{ \chi_{a}(x), \bar{\lambda}_{\bar{b}}(y) \right\} =
- i~(c_{4} + i c_{3})~\sigma^{\mu}_{a\bar{b}}~\partial_{\mu}D_{m}(x-y)
\label{CCR-V}
\eea
and all other (anti)commutators are zero. 

Finally we give the gauge structure of the vector multiplet. We need 
three ghost superfields
$U, \tilde{U}, H$
(the first two fermionic and the last one bosonic) which are analogues
of the usual ghost fields of the standard model \cite{Sc}. We take
these fields to be chiral
\be
{\cal D}_{a}~U = 0 \quad {\cal D}_{a}~\tilde{U} = 0 \quad {\cal D}_{a}~H = 0.
\ee

In components we have
\bea
H(x,\theta,\bar{\theta}) = h(x) 
+ 2~\bar{\theta} \bar{\psi}(x)
+ i~(\theta \sigma^{\mu} \bar{\theta})~\partial_{\mu}h(x)
+ \bar{\theta}^{2}~f(x)
- i~\bar{\theta}^{2}~\theta \sigma^{\mu} \partial_{\mu}\bar{\psi}(x)
\nonumber \\
+ {m^{2} \over 4}~\theta^{2} \bar{\theta}^{2}~h(x)
\label{chiral}
\eea
where
$h, f$
and are scalar and
$\psi$
is a Dirac field. Analogously we have:
\bea
U(x,\theta,\bar{\theta}) = u(x) 
+ 2i~\bar{\theta} \bar{\zeta}(x)
+ i~(\theta \sigma^{\mu} \bar{\theta})~\partial_{\mu}u
+ \bar{\theta}^{2}~g(x)
+ \bar{\theta}^{2}~\theta \sigma^{\mu} \partial_{\mu}\bar{\zeta}(x)
+ {m^{2} \over 4}~\theta^{2} \bar{\theta}^{2}~u(x)
\label{U}
\eea
and, respectively
\bea
\tilde{U}(x,\theta,\bar{\theta}) = \tilde{u}(x) 
- 2i~\bar{\theta} \bar{\tilde{\zeta}}(x)
+ i~(\theta \sigma^{\mu} \bar{\theta})~\partial_{\mu}\tilde{u}
+ \bar{\theta}^{2}~\tilde{g}(x)
- \bar{\theta}^{2}~\theta \sigma^{\mu} \partial_{\mu}\bar{\tilde{\zeta}}(x)
+ {m^{2} \over 4}~\theta^{2} \bar{\theta}^{2}~\tilde{u}(x)
\label{tildeU}
\eea
where
$u, g$
(resp.
$\tilde{u}, \tilde{g}$)
are scalar fields of Fermi statistics
and
$\zeta$
(resp.
$\tilde{\zeta}$ )
are Dirac fields of Bose statistics. The superfields
$U, \tilde{U}, H$
are of the same mass $m$ as the vector multiplet.

Now we give the definition of the gauge charge. In ordinary quantum gauge 
theory, one gauges away the unphysical degrees of freedom of a vector field 
$v_{\mu}$
using ghost fields. Suppose that the vector field is of positive mass $m$; 
then one enlarges the Hilbert space with three ghost fields
$v,~\tilde{v},~\phi$
such that:
(a) All three are scalar fields;
(b) All them have the same mass $m$ as the vector field;
(c) The Hermiticity properties with respect to the indefinite sesqui-linear 
form are;
\be
\phi^{\dagger} = \phi, \qquad v^{\dagger} = v, \qquad
\tilde{v}^{\dagger} = - \tilde{v};
\ee
(d) The last two ones 
$v, ~\tilde{v}$
are fermionic and 
$\phi$
is bosonic;
(e) The commutation relations for the ghost fields are:
\bea
~[ \phi(x), \phi(y) ] = - i~D_{m}(x - y), \qquad
~\{ v(x), \tilde{v}(y) \} = - i~D_{m}(x - y)
\eea
and the rest of the (anti)commutators are zero.
Then one introduces the {\it gauge charge} $Q$ according to:
\bea
Q \Omega = 0 \quad Q^{\dagger} = Q
\nonumber \\
~[ Q, v_{\mu} ] = i \partial_{\mu}v, \qquad [ Q, \phi ] = i~m~v
\nonumber \\
~\{ Q, v \} = 0, \qquad 
\{ Q, \tilde{v} \} = - i~(\partial^{\mu}v_{\mu} + m~\phi).
\label{gh-charge}
\eea

It can be proved that this gauge charge is well defined by these relations
i.e. it is compatible with the (anti)commutation relations.
Moreover one has
$Q^{2} = 0$
so the factor space
$Ker(Q)/Im(Q)$
makes sense; it can be proved that this is the physical space of an ensemble
of identical particles of spin $1$. For details see \cite{Sc}, \cite{Gr1}.

In \cite{GS2} we have generalized this structure for the vector multiplet:
\be
Q \Omega = 0 \quad Q^{\dagger} = Q
\label{gauge1}
\ee

and
\bea
~[ Q, V ] = U - U^{\dagger}
\nonumber \\
~\{ Q, U \} = 0.
\nonumber \\
\{ Q, \tilde{U} \} = - {1\over 16}~{\cal D}^{2}~\bar{\cal D}^{2}V - i~m~H
\nonumber \\
~[ Q, H] = i~m~U.
\label{gauge2}
\eea

Let us note in particular
\be
~[ Q, D^{\prime} ] = 0. 
\ee

These commutators and anticommutators with $Q$ define the gauge
variation
$d_{Q}$
of the operators as the graded bracket with $Q$.
It is interesting to translate (\ref{gauge2}) in terms of the component 
fields of the multiplet. One notices in this way that the fields
\be
v_{\mu}, u + u^{\dagger}, \tilde{u} - \tilde{u}^{\dagger}, h + h^{\dagger}.
\label{gh-structure}
\ee
have the same gauge structure as (\ref{gh-charge}). However, let us
emphasize that our model will be different from the usual standard model
because the number of ghost fields is doubled.

Now we consider a generic form for the one-particle states from the
Hilbert space. It can be written in a supersymmetric way as follows:
\bea
\Psi = \int f_{1}(X) V(X) + \int f_{2}(X) U(X) + 
\int f_{3}(X) U(X)^{\dagger} + \int f_{4}(X) \tilde{U}(X) 
\nonumber \\
+ \int f_{5}(X) \tilde{U}(X)^{\dagger} + \int f_{6}(X) H(X) 
+ \int f_{7}(X) H(X)^{\dagger} 
\label{test}
\eea
with
$f_{j}, j = 1, \dots, 7$
supersymmetric test functions. The integration is over
$d^{4}x d^{2}\theta d^{2}\bar{\theta}$.
Let us note that the writing (\ref{test}) is unique {\it iff} we require 
\be
P_{2}f_{j} = 0, \quad j = 2, 4, 6
\qquad
P_{1}f_{j} = 0, \quad j = 3, 5, 7
\ee
because of the chirality condition. Indeed we have for instance
\bea
\int f_{2}(X) U(X) = \int (P_{2}f_{2})(X) U(X)
\eea
and similar relations for the other chiral fields. Also because of (\ref{kg})
we can suppose that all these test functions verify Klein-Gordon equation.

If we impose the condition
$\Psi \in Ker(Q)$
i.e.
$Q \Psi = 0$
then we immediately find
\be
P_{1}f_{1} = P_{2}f_{1} = 0
\label{transv}
\ee
\be
f_{4} = 0, \quad f_{5} = 0.
\ee
Now the generic expression of an element from
$Im(Q)$
is 
\bea
\Psi^{\prime} = \int (g_{1} + i m g_{6})(X) U(X) 
+ \int (- g_{1} + i m g_{7})(X) U(X)^{\dagger} 
\nonumber \\
- {1 \over 16}\int \bar{\cal D}^{2} {\cal D}^{2})g_{4}
+ {\cal D}^{2} \bar{\cal D}^{2}g_{5})(X) V(X)
- i m \int g_{4}(X) H(X) + i m \int g_{5}(X) H(X)^{\dagger}. 
\label{im-q}
\eea

If we take the test functions
$g_{j}, j = 1, \dots, 7$
convenient, then we can arrange such that
\be
\Psi - \Psi^{\prime} = \int f_{1}(X) V(X)
\ee
where
$f_{1}$
is restricted only by (\ref{transv}). Moreover one can see that in
every equivalence classes from
$Ker(Q)/Im(Q) \cap {\cal H}^{(1)}$
there exists one and only one element of transverse form; so the
equivalence classes are indexed by supersymmetric transversal functions. 
Equivalently, only the transversal part of the vector superfield is producing
physical states. This statement is the rigorous form of the so-called
{\it Wess-Zumino gauge}. 
\begin{rem}
The preceding argument depends essentially on the positivity of the mass: 
for a null mass we do not have anymore the projection operators
$P_{j}$.
Also one should generalize this argument to multi-particle states. This
is a difficult technical problem.
\end{rem}
From the preceding analysis it follows that we have to impose the 
positivity of the scalar product only in the sector generated by the 
fields appearing in the superfield
$D^{\prime} = - {m^{2} \over 2} V_{0}$
i.e.
$d^{\prime}, \lambda^{\prime}$
and
the transversal part of
$v_{\mu}$.
If we do this by going through the proof of theorem 3.8 of \cite{GS2},
we get only the restriction
\be
c_{2} \leq 0.
\ee
One can see this directly from
\bea
~[ d^{\prime}(x), d^{\prime}(y) ] = - {im^{4}\over 8}~
(c_{1} - 2 {c_{2} \over m^{2}})~D_{m}(x-y)
\nonumber \\
~[ v_{\mu}^{\prime}(x), v_{\rho}^{\prime}(y) ] = 
i~\left(c_{1} - 2 {c_{2} \over m^{2}}\right)~
(\partial_{\mu}\partial_{\rho} + m^{2} g_{\mu\rho})~D_{m}(x-y)
\nonumber \\
\left\{ \lambda_{a}^{\prime}(x), \lambda_{b}^{\prime}(y) \right\} 
= - {m^{2}\over 2}(c_{4} - i c_{3})~\epsilon_{ab}~D_{m}(x-y),
\nonumber \\
\left\{ \lambda_{a}^{\prime}(x), \bar{\lambda}_{\bar{b}}^{\prime}(y) \right\} =
- 2~c_{2}~\sigma^{\mu}_{a\bar{b}}~\partial_{\mu}D_{m}(x-y).
\label{CCR-V-prime}
\eea

If we want the field
$v_{\mu}$
to give a renormalizable theory then we should also put
\be
c_{1} = 0
\ee
in such a way that the causal commutator of
$v_{\mu}$
does not have derivatives in the right hand side (\ref{CCR-V}).
In particular, the solution
$D(X_{1};X_{2}) = - D_{2}(X_{1};X_{2})$
suggested by the supersymmetry literature is acceptable from the point of
view of positivity.

We also mention that the preceding analysis is compatible with the
following SUSY-scalar product
\be
<f_{1}, f_{2}> \equiv 
- \int f_{1}(X_{1})^{\dagger} D_{2}(X_{1};X_{2}) f_{2}(X_{2})
\ee
like is suggested in \cite{Co}.
\section{A Massive Supersymmetric Gauge Invariant Coupling\label{lagr}}

We consider the following interaction Lagrangian:
\be
T = \sum_{j=1}^{7} T^{(j)}
\ee
with
\bea
T^{(1)} \equiv f^{(1)}_{jkl} \left[ :V_{j} ({\cal D}^{a}V_{k})
(\bar{\cal D}^{2} {\cal D}_{a}V_{k}): - H. c. \right] 
\nonumber \\
T^{(2)} \equiv f^{(2)}_{jkl} :V_{j}~(U_{k} + U_{k}^{\dagger})
(\tilde{U}_{l} + \tilde{U}_{l}^{\dagger}):
\nonumber \\
T^{(3)} = f^{(3)}_{jkl}~:(H_{j} + H_{j}^{\dagger})~
(U_{k} - U_{k}^{\dagger}) (\tilde{U}_{l} + \tilde{U}^{\dagger}_{l}):
\nonumber \\
T^{(4)} = f^{(4)}_{jkl}~:(H_{j} + H_{j}^{\dagger})~
(H_{k} - H_{k}^{\dagger}) V_{l}:
\nonumber \\
T^{(5)} = f^{(5)}_{jkl}~:(H_{j} + H_{j}^{\dagger})~V_{k}~D^{\prime}_{l}
\nonumber \\
T^{(6)} = f^{(6)}_{jkl}~:(H_{j} + H_{j}^{\dagger})~V_{k}~V_{l}:
\nonumber \\
T^{(7)} = f^{(7)}_{jkl}~:(H_{j} + H_{j}^{\dagger})~
(H_{k} - H_{k}^{\dagger}) D^{\prime}_{l}:
\label{lagrange}
\eea
where 
$f^{(j)}_{jkl}, j = 1, \dots, 7$
are some constants to be determined later. The last term is new by comparison
to \cite{GS2}. Because of the identity
\be
:(H_{j} + H_{j}^{\dagger})~(H_{k} - H_{k}^{\dagger}) D^{\prime}_{l}:
+ :(H_{j} - H_{j}^{\dagger})~(H_{k} + H_{k}^{\dagger}) D^{\prime}_{l}:
= {\rm SUSY-divergence}
\ee
we can suppose that
\be
f^{(7)}_{jkl} = - f^{(7)}_{kjl}.
\label{s1}
\ee
We can also take
\be
f^{(6)}_{jkl} = f^{(6)}_{jlk}.
\label{s2}
\ee
As discussed in \cite{GS2}, the list (\ref{lagrange}) is suggested by
the fact that after integrating out the Grassmann variables one gets
coupling terms of the same form as in ordinary gauge theory.

We impose the supersymmetric gauge invariance condition (\ref{gauge-susy})
from the Introduction. Noticing that
$d_{Q} D^{\prime}  = 0$
we have
\bea
d_{Q} T^{(1)} = - 32 f^{(1)}_{jkl}~:(U_{j}  + U_{j}^{\dagger})~V_{k}
D^{\prime}_{l} + {\rm SUSY-div}
\nonumber \\
d_{Q} T^{(2)} = f^{(2)}_{jkl}~:(U_{j} - U_{j}^{\dagger})~
(U_{k} + U_{k}^{\dagger})~(\tilde{U}_{l} + \tilde{U}_{l}^{\dagger}):
\nonumber \\
+ 2~f^{(2)}_{jkl}~:V_{j}~(U_{k} + U_{k}^{\dagger})~D^{\prime}_{l}
\nonumber \\
+ f^{(2)}_{jkl}~m_{l}^{2} :V_{j}~(U_{k} + U_{k}^{\dagger})~V_{l}
\nonumber \\
+ i~f^{(2)}_{jkl}~m_{l}:V_{j}~(U_{k} + U_{k}^{\dagger})~
(H_{l} - H_{l}^{\dagger}):
\nonumber \\
d_{Q} T^{(3)} = i~f^{(3)}_{jkl}~m_{j}:(U_{j} + U_{j}^{\dagger})~
(U_{k} - U_{k}^{\dagger})~(\tilde{U}_{l} + \tilde{U}_{l}^{\dagger}):
\nonumber \\
+ 2~f^{(3)}_{jkl}~:(H_{j} + H_{j}^{\dagger})~(U_{k} - U_{k}^{\dagger})~
D^{\prime}_{l}
\nonumber \\
+ f^{(3)}_{jkl}~m_{l}^{2} :(H_{j} + H_{j}^{\dagger})~
(U_{k} - U_{k}^{\dagger})~V_{l}
\nonumber \\
+ i~f^{(3)}_{jkl}~m_{l}:(H_{j} + H_{j}^{\dagger})~
(U_{k} - U_{k}^{\dagger})~(H_{l} - H_{l}^{\dagger}):
\nonumber \\
d_{Q} T^{(4)} = i~f^{(4)}_{jkl}~m_{j}:(U_{j} + U_{j}^{\dagger})~
(H_{k} - H_{k}^{\dagger})~V_{l}:
\nonumber \\
+ i~f^{(4)}_{jkl}~m_{k}:(H_{j} + H_{j}^{\dagger})~(U_{k} - U_{k}^{\dagger})~
V_{l}
\nonumber \\
+ f^{(4)}_{jkl}~:(H_{j} + H_{j}^{\dagger})~(H_{k} - H_{k}^{\dagger})~
(U_{l} - U_{l}^{\dagger})
\nonumber \\
d_{Q} T^{(5)} = i~f^{(5)}_{jkl}~m_{j}:(U_{j} + U_{j}^{\dagger})~
V_{k}~D^{\prime}_{l}:
\nonumber \\
+ f^{(5)}_{jkl}:(H_{j} + H_{j}^{\dagger})~(U_{k} - U_{k}^{\dagger})~
D^{\prime}_{l}:
\nonumber \\
d_{Q} T^{(6)} = i~f^{(6)}_{jkl}~m_{j}:(U_{j} + U_{j}^{\dagger})~
V_{k}~V_{l}:
\nonumber \\
+ 2~f^{(6)}_{jkl}:(H_{j} + H_{j}^{\dagger})~(U_{k} - U_{k}^{\dagger})~
V_{l}:
\nonumber \\
d_{Q} T^{(7)} = 2 i~f^{(7)}_{jkl}~m_{k}:(H_{j} + H_{j}^{\dagger})~
(U_{k} - U_{k}^{\dagger})~D^{\prime}_{l}: +  {\rm SUSY-div}
\eea

To obtain the last expression one must use the identity
\be
:(U_{j} + U_{j}^{\dagger})~(H_{k} - H_{k}^{\dagger}) D^{\prime}_{l}:
+ :(U_{j} - U_{j}^{\dagger})~(H_{k} + H_{k}^{\dagger}) D^{\prime}_{l}:
= {\rm SUSY-divergence}.
\ee

The gauge invariance condition (\ref{gauge-susy}) now leads to the
following system:
\bea
f^{(2)}_{kjl}~m_{l}^{2} + i~f^{(6)}_{jkl}~m_{j} = - ( k \leftrightarrow l)
\nonumber \\
- 32 f^{(1)}_{jkl} + 2~f^{(2)}_{kjl} + i~f^{(5)}_{jkl}~m_{j} = 0
\nonumber \\
f^{(2)}_{jkl} - i~f^{(3)}_{kjl}~m_{k} = 0
\nonumber \\
f^{(2)}_{kjl} m_{l} + f^{(4)}_{jlk} m_{j} = 0
\nonumber \\
2 f^{(3)}_{jkl} + f^{(5)}_{jkl} + 2 i~f^{(7)}_{jkl}~m_{k} = 0
\nonumber \\
f^{(3)}_{jkl}~m_{l}^{2} + i~f^{(4)}_{jkl}~m_{k} + 2 f^{(67)}_{jkl} = 0
\nonumber \\
i~f^{(3)}_{jkl} m_{l} + f^{(4)}_{jlk} = 0
\label{f-system}
\eea

The solution of this system is determined by the expressions
$f^{(1)}_{jkl}$
and
$f^{(2)}_{jkl}$;
explicitly
\bea
f^{(3)}_{jkl} =  - {i \over m_{j}} f^{(2)}_{kjl}
\nonumber \\
f^{(4)}_{jkl} = - {m_{k} \over m_{j}}~f^{(2)}_{ljk}
\nonumber \\
f^{(5)}_{jkl} = {2 i\over m_{j}}~( f^{(2)}_{kjl} - 16 f^{(1)}_{jkl} )
\nonumber \\
f^{(6)}_{jkl} = {i\over 2 m_{j}} 
(m_{k}^{2}~f^{(2)}_{ljk} + m_{l}^{2}~f^{(2)}_{kjl})
\nonumber \\
f^{(7)}_{jkl} =  {16 \over m_{j} m_{k}} f^{(1)}_{jkl}
\label{f's}
\eea
where we also have
\be
f^{(1)}_{jkl} = - f^{(1)}_{kjl}.
\ee

If one computes the corresponding expression $t$
(see the Introduction) by integrating out the Grassmann variables one gets, 
up to finite renormalizations, the following expressions:
\bea
\int d\theta^{2} d\bar{\theta}^{2} T^{(1)} =
4 i~f^{(1)}_{jkl} :v^{\mu}_{j} v^{\nu}_{k}~
(\partial_{\nu}v_{l\mu} - \partial_{\mu}v_{l\nu}): + \cdots
\nonumber \\
\int d\theta^{2} d\bar{\theta}^{2} T^{(2)} =
{i\over 2}~f^{(2)}_{jkl} :v^{\mu}_{j} [ (u_{k} + u_{k}^{\dagger})~
\partial_{\mu}(\tilde{u}_{l} - \tilde{u}_{l}^{\dagger})
+ \partial_{\mu}(u_{k} - u_{k}^{\dagger})~
(\tilde{u}_{l} + \tilde{u}_{l}^{\dagger})]: + \cdots
\nonumber \\
\int d\theta^{2} d\bar{\theta}^{2} T^{(3)} =
{1\over 4}~f^{(3)}_{jkl} \Bigl\{
( m^{2}_{j} + m^{2}_{k} + m^{2}_{l}) :(h_{j} + h_{j}^{\dagger})~
(u_{k} - u_{k}^{\dagger})~(\tilde{u}_{l} - \tilde{u}_{l}^{\dagger}): 
\nonumber \\
- ( m^{2}_{j} + m^{2}_{k} - m^{2}_{l}) :(h_{j} - h_{j}^{\dagger})~
(u_{k} + u_{k}^{\dagger})~(\tilde{u}_{l} + \tilde{u}_{l}^{\dagger}): 
\nonumber \\
- ( m^{2}_{k} + m^{2}_{l} - m^{2}_{j}) :[ (h_{j} + h_{j}^{\dagger})~
(u_{k} + u_{k}^{\dagger}) + (h_{j} - h_{j}^{\dagger})~
(u_{k} - u_{k}^{\dagger})]~(\tilde{u}_{l} - \tilde{u}_{l}^{\dagger}):
\Bigl\} + \cdots
\nonumber \\
\int d\theta^{2} d\bar{\theta}^{2} T^{(4)} =
{i\over 2}~f^{(4)}_{jkl} :[ 
(h_{j} + h_{j}^{\dagger})~\partial_{\mu}(h_{k} + h_{k}^{\dagger})
+ (h_{j} - h_{j}^{\dagger})~\partial_{\mu}(h_{k} - h_{k}^{\dagger})]
v^{\mu}_{l}: + \cdots
\nonumber \\
\int d\theta^{2} d\bar{\theta}^{2} T^{(5)} =
- {1\over 4}~f^{(5)}_{jkl} m_{l}^{2}
:(h_{j} + h_{j}^{\dagger})~v_{k}^{\mu}~v_{l\mu}^{\prime}: + \cdots
\nonumber \\
\int d\theta^{2} d\bar{\theta}^{2} T^{(6)} =
{1\over 2}~f^{(6)}_{jkl}~:(h_{j} + h_{j}^{\dagger})~v_{k}^{\mu}~v_{l\mu}: 
+ \cdots
\nonumber \\
\int d\theta^{2} d\bar{\theta}^{2} T^{(7)} =
{i\over 2}~f^{(7)}_{jkl}~( m^{2}_{j} - m^{2}_{k} + m^{2}_{l}) 
\nonumber \\ \times
:[(h_{j} + h_{j}^{\dagger})~\partial_{\mu}(h_{k} + h_{k}^{\dagger}) 
+ (h_{j} - h_{j}^{\dagger})~\partial_{\mu}(h_{k} - h_{k}^{\dagger})]~
v_{l}^{\mu}: + \cdots
\eea
where by 
$\cdots$
we mean terms containing the superpartners from the corresponding multiplets.

It seems encouraging that the first term is in agreement with the usual
gauge coupling of the vector fields in Yang-Mills theory.
After Grassmann integration one finds out that
$T^{(5)}$
is producing a non-renormalizable expression. So in the following
we consider the case
\be
f^{(5)}_{jkl} = 0.
\ee
In this case one can re-express everything in terms of 
\be
f_{jkl} \equiv f^{(1)}_{jkl}
\ee
which verifies
\be
f_{jkl} = - f_{kjl}
\ee
as follows:
\bea
f^{(2)}_{jkl} = 16~f_{kjl}, \qquad
%\nonumber \\
f^{(3)}_{jkl} =  {- 16 i \over m_{j}} f_{jkl}, \qquad
%\nonumber \\
f^{(4)}_{jkl} = - { 16 m_{k} \over m_{j}}~f_{jlk}
\nonumber \\
f^{(6)}_{jkl} = {8 i\over m_{j}} 
(m_{k}^{2}~f_{jlk} + m_{l}^{2}~f_{jkl}), \qquad
%\nonumber \\
f^{(7)}_{jkl} =  {16 \over m_{j} m_{k}} f_{jkl}.
\label{f's1}
\eea

This is our massive supersymmetric quantum gauge model. Let us also
remark that the doubling of the number of ghost fields cannot be
avoided. The only way of halving this number would be to consider
that the super-ghost multiplets are of Wess-Zumino type. However,
this assumption together with (\ref{gauge2}) leads to a contradiction. 

A strange solution corresponding to the case
$f^{(5)}_{jkl} \not= 0$
was already discovered in \cite{GS2}.

\section{Conclusions}

Our main result is the existence of a physically interesting massive
super-gauge theory satisfying the strong supersymmetric gauge condition
(\ref{gauge-susy}). This condition is the true supersymmetric extension 
of first order perturbative gauge invariance \cite{Sc}, \cite{Gr1}. It 
expresses gauge invariance in terms of the chronological products using 
the gauge structure of the free asymptotic fields only, whereas other
formulations of gauge invariance (classical or quantum) deal with interacting
fields. The usefulness of (\ref{gauge-susy}) is evident from the fact that 
it determines the theory essentially unique under the renormalizable theories 
of the form (\ref{lagrange}). 

In ordinary gauge invariance we know that
$f_{jkl}$
has to be totally antisymmetric. Assuming this we find the coupling
\bea
T = f_{jkl} \Bigl[ :V_{j} ({\cal D}^{a}V_{k})
(\bar{\cal D}^{2} {\cal D}_{a}V_{k}): - H. c.
\nonumber \\
+ 16 :V_{j}~(U_{k} + U_{k}^{\dagger})
(\tilde{U}_{l} + \tilde{U}_{l}^{\dagger}):
- {16 i \over m_{j}}~:(H_{j} + H_{j}^{\dagger})~
(U_{k} - U_{k}^{\dagger}) (\tilde{U}_{l} + \tilde{U}^{\dagger}_{l}):
\nonumber \\
+ {16 m_{k} \over m_{j}}~:(H_{j} + H_{j}^{\dagger})~
(H_{k} - H_{k}^{\dagger}) V_{l}:
%\nonumber \\
+ {8 i\over m_{j}} (m_{l}^{2} - m_{k}^{2})~:(H_{j} + H_{j}^{\dagger})~
V_{k}~V_{l}:
\nonumber \\
- {16 \over m_{j} m_{k}}~:(H_{j} + H_{j}^{\dagger})~
(H_{k} - H_{k}^{\dagger}) D^{\prime}_{l}: \Bigl]
\label{lag}
\eea
or, after Grassmann integration
\bea
t = \int d\theta^{2} d\bar{\theta}^{2} T = 4 i~f_{jkl}
\Bigl\{ 2 :v^{\mu}_{j} v^{\nu}_{k}~
\partial_{\nu}v_{l\mu}:
\nonumber \\
- 2~:v^{\mu}_{j} [ (u_{k} + u_{k}^{\dagger})~
\partial_{\mu}(\tilde{u}_{l} - \tilde{u}_{l}^{\dagger})
+ \partial_{\mu}(u_{k} - u_{k}^{\dagger})~
(\tilde{u}_{l} + \tilde{u}_{l}^{\dagger})]:
\nonumber \\
- {1 \over m_{j}}~\Bigl[
( m^{2}_{j} + m^{2}_{k} + m^{2}_{l}) :(h_{j} + h_{j}^{\dagger})~
(u_{k} - u_{k}^{\dagger})~(\tilde{u}_{l} - \tilde{u}_{l}^{\dagger}): 
\nonumber \\
- ( m^{2}_{j} + m^{2}_{k} - m^{2}_{l}) :(h_{j} - h_{j}^{\dagger})~
(u_{k} + u_{k}^{\dagger})~(\tilde{u}_{l} + \tilde{u}_{l}^{\dagger}): 
\nonumber \\
- ( m^{2}_{k} + m^{2}_{l} - m^{2}_{j}) :\Bigl( (h_{j} + h_{j}^{\dagger})~
(u_{k} + u_{k}^{\dagger}) + (h_{j} - h_{j}^{\dagger})~
(u_{k} - u_{k}^{\dagger})\Bigl)~(\tilde{u}_{l} - \tilde{u}_{l}^{\dagger}):
\Bigl]
\nonumber \\
+ {1 \over m_{j} m_{k}}~[ ( 3 m^{2}_{k} - m^{2}_{j} - m^{2}_{l}) 
:(h_{j} + h_{j}^{\dagger})~\partial_{\mu}(h_{k} + h_{k}^{\dagger})
\nonumber \\
+ ( m^{2}_{k} + m^{2}_{j} - m^{2}_{l})~ 
\partial_{\mu}(h_{j} - h_{j}^{\dagger})~(h_{k} - h_{k}^{\dagger})]~
v_{l}^{\mu}:
\nonumber \\
+ {1 \over m_{j}}~(m_{k}^{2} - m_{l}^{2})
:(h_{j} + h_{j}^{\dagger})~v_{k}^{\mu}~v_{l\mu}: \Big\} + \cdots
\eea
where by 
$\cdots$
we mean terms containing the superpartners from the corresponding multiplets.

We note for instance, that the third term of the preceding expression
is absent in the ordinary Yang-Mills coupling.
On the other hand the theory is not yet completely specified because there
are still some free parameters in the commutations rules of $V$ and the
ghost fields. Furthermore, in second order gauge invariance one expects the 
necessity of a Higgs superfield as in ordinary massive gauge theory. This
will be investigated in a future work.
%\newpage

\end{document}